\documentclass[runningheads]{llncs}
\usepackage[T1]{fontenc}
\usepackage{graphicx}
\usepackage{booktabs}
\usepackage{amssymb}
\usepackage{subcaption}
\usepackage{amsmath}
\usepackage{tabularx}
\usepackage{array}

\usepackage{xspace}

\begin{document}
\title{The PM-EdgeMap: Towards Real-Time Process Mining on the Edge-Cloud Continuum}

\author{%
Hendrik Reiter\orcidID{0009-0003-8544-0012} \inst{1}
\and Christian Imenkamp\orcidID{0009-0007-4295-1268} \inst{2}
\and Olaf Landsiedel\orcidID{0000-0001-6432-300X} \inst{1}
\and Andrea Maldonado\orcidID{0009-0009-8978-502X} \inst{3}
\and Patrick Rathje\orcidID{0000-0003-3718-7115} \inst{1}
\and Wilhelm Hasselbring\orcidID{0000-0001-6625-4335} \inst{1}
}
\authorrunning{H. Reiter et al.}
\titlerunning{Real-Time Process Mining on the Edge-Cloud Continuum}
%
\institute{Kiel University, Christian-Albrechts-Platz 4, 24118 Kiel, Germany 
\email{\{hendrik.reiter, patrick.rathje, olaf.landsiedel, hasselbring\}@email.uni-kiel.de},\\
\and
University of Bayreuth, Germany
\email{christian.imenkamp@uni-bayreuth.de}\\
\and
School of Engineering and Design, Technical University of Munich, Germany
\email{andrea.maldonado@tum.de}}

\maketitle              
\begin{abstract}
Smart factories are evolving into Cyber-Physical Systems (CPS), demanding increased autonomy. This necessitates real-time decision making, facilitated by insights derived from sensor data. Process mining offers a valuable approach to gain such insights and guide actions. The edge computing paradigm supports this real-time requirement by enabling network communication between sensors and leveraging nearby computing resources. This paper investigates the implications of performing real-time process mining algorithms on the edge. Within this paper, we first propose a formalism to describe relevant datasets and the computing topology. We then evaluate the edge computing approach through a case study involving an edge-based conformance checking algorithm. The results demonstrate the feasibility and benefits of edge-based real-time process mining for enhanced autonomous control in smart factories.
\end{abstract}

\section{Introduction}

The ongoing evolution of industrial automation towards highly autonomous systems within the Industrial Internet of Things (IIoT) represents a significant paradigm shift in manufacturing and operational management~\cite{Babayigit2024}. This transformation is largely facilitated by the increasing integration of smart machines and advanced robotics, which, equipped with sophisticated sensing and actuation capabilities, function as core components of Cyber-Physical Systems (CPS)~\cite{baheti}. These systems bridge the physical and digital domains, enabling intricate control and interaction within complex industrial environments.
A fundamental characteristic of these CPS is the pervasive generation of data by numerous sensors. This sensor data exhibits properties aligned with Big Data challenges~\cite{Qi2023}, specifically concerning its volume, velocity, variety, and veracity. For intelligent decision-making and the effective realization of autonomous control, this data requires processing not only in real-time but also with a high degree of accuracy. Traditional, centralized data processing architectures frequently encounter limitations in addressing these demanding requirements, particularly given the scale and rate of data generation.

To address these challenges, the edge computing paradigm has emerged as an architectural solution for contemporary IIoT applications~\cite{Chalapathi2021} such as 
real-time machine data analytics, image classification, or real-time anomaly detection~\cite{Bayar2023}. Edge computing~\cite{Satyanarayanan2017} involves distributing computational resources closer to the data sources, leveraging a network of heterogeneous, interconnected devices. This distributed architecture offers distinct advantages in fulfilling real-time processing demands by reducing data latency, optimizing network bandwidth utilization by minimizing the transfer of raw data to central repositories, and enhancing data privacy through localized data management. 
By providing immediate computational capabilities at or near the point of data generation, edge computing~\cite{Satyanarayanan2017} aims to support the real-time requirements of autonomous systems. 
Process mining~\cite{vanderAalst2022} is a discipline for gaining insights from event data related to operational processes with techniques such as process discovery or conformance checking. Moreover, process mining has expanded to be performed on event streams~\cite{Burattin2022}, which processes event data as it is generated, enabling near-real-time analysis.

Considering the intersection between IoT factory automation and Process mining, a notable research gap exists when performing streaming process mining on edge computing topologies. Both disciplines can benefit from a synergistic approach, particularly within the context of IIoT. Conventionally, process mining has been conceptualized primarily as a centralized activity, performing computations on a single, aggregated event log within a central computing instance. In contrast, edge computing operates on distributed and localized data streams, necessitating computations to be performed with only a subset of the complete system data. This architectural divergence between traditional process mining and the distributed characteristics of edge computing presents an unsolved challenge and a significant avenue for research.
This paper addresses this research gap by investigating the implications of incorporating edge computing in the context of process mining. 
We present the Process Mining Edge Map (PM-EdgeMap), a formalization of process mining algorithms and their quality attributes in the realm of edge computing. Thereby, the PM-EdgeMap defines how algorithms interact with edge topologies and their distributed event streams. In summary, the paper contributes the following:

\begin{enumerate}
    \item The PM-EdgeMap: A formalization of edge computing in the realm of process mining, their quality measures, and distributed event streams in the realm of process mining.
    \item A prototype of an edge conformance checking algorithm, designed to demonstrate the suitability of the PM-EdgeMap.
\end{enumerate}

\section{Motivation}

Consider a modern smart factory producing high-precision automotive components using a network of interconnected robotic workstations.
Each workstation is equipped with sensors that continuously log events such as task completions or inter-robot handovers.
In such a setting, maintaining tight control over the production process is critical, not only to meet quality standards but also to ensure safety and minimize downtime.
A key challenge arises when deviations from the expected process behavior occur, such as unexpected delays, or incorrect task sequences.
This is where real-time process mining analysis becomes crucial in tasks such as conformance checking, which detects deviations from the defined production model.
Traditional process mining approaches first send the data to a central computing instance. This takes time and requires significant network bandwidth. And in turn, may violate the real-time constraints for safety-critical applications, particularly when video streams or sensor time series have to be transmitted to the central computing instance.
However, by deploying these tasks directly on devices close to the machines, i.e., on the edge, these edge devices can detect anomalies immediately and adapt control strategies, without the need to transfer large amounts of raw data to a central server.
For example, suppose an edge-based conformance checking algorithm detects that a robotic arm skipped a critical assembly step. In that case, the system can halt the workflow or reroute products before defects propagate downstream.
In general, process mining can benefit from edge computing by enabling real-time analysis of several process-related tasks mentioned in the IoT-meets-BPM Manifesto~\cite{Janiesch2020}, such as online conformance checking, resource utilization optimization, or evaluation of the quality of task execution. Here, real-time analyses provide operational support that allows interactions with the process while it is still running. In this paper, we will argue that process mining can benefit from edge computing by (1) enabling real-time process analysis which supports autonomous control in IIoT systems, (2) utilizing fewer cloud resources which saves computing costs, and (3) more event data can be processed which allows more fine-grained process mining results.
\section{Edge Computing}
\label{sec:edge}

Edge computing~\cite{Satyanarayanan2017} represents a distributed computing paradigm that strategically positions computational capabilities closer to the data sources, leveraging a network of heterogeneous, interconnected devices.
Thereby it contrasts traditional cloud computing models, in which data is primarily transmitted to and processed within central data centers. This section characterizes edge computing by describing its primary motivations, the diverse technical infrastructure it employs, and the critical quality attributes that define its performance and utility.


\begin{table}[h] 
    \centering
    \caption{Results from the literature concerning edge computing in the categories main motivations, edge computing topologies, and quality attributes}
    \begingroup 
    \setlength{\tabcolsep}{3pt} 
    \begin{tabular}{|p{0.32\textwidth}|p{0.32\textwidth}|p{0.32\textwidth}|}
        \hline
        \textbf{Objectives}~\cite{Satyanarayanan2017} & \textbf{Topologies}~\cite{Mahmud2017,AlDulaimy2024} & \textbf{Quality Attributes}~\cite{Ashouri2021} \\
        \hline
        Reduced Latency & Mobile Cloud Comp. & Time Behavior \\
        Low Bandwidth Utilization & Mobile Edge Comp. & Resource Utilization \\
        Privacy \& Security & Cloudlet Computing & Load Capacity \\
        Scalability \& Reliability & & Scalability \\
        \hline
    \end{tabular}
    \endgroup
    \label{tab:motivations}
\end{table}



\textbf{Edge Computing Objectives.}
The adoption of edge computing is driven by four key objectives: reduced \textit{network latency}, improved \textit{bandwidth utilization}, enhanced \textit{privacy and security}, and improved \textit{scalability and reliability}~\cite{Satyanarayanan2017}, as found in literature and summarized in Table~\ref{tab:motivations}.

Locating computing resources in closer geographical and network proximity to data generation points reduces \textit{latency} due to the physical distance within the network. This proximity minimizes data transmission delays, leading to lower latency for data processing and subsequent response generation.
Edge computing enhances \textit{bandwidth utilization} by reducing the need to transfer all raw data across wide area networks to centralized cloud infrastructures. Techniques such as data filtering, aggregation, and batching can be applied to the edge data before transmitting. For instance, in video surveillance, instead of streaming raw video footage, an edge device can classify objects or detect events locally and transmit only metadata or alerts.
By design, edge computing inherently supports improved \textit{data privacy} as processing data closer to its origin means that sensitive or raw information may not need to be transferred beyond the local device or network segment. Furthermore, through local abstraction and aggregation, edge nodes can derive insights or aggregated results, transmitting only this derived, non-sensitive information. This minimizes the exposure of private data to wider networks and central cloud systems, thereby reducing potential attack surfaces and enhancing overall data security.
The distributed nature of edge computing contributes to enhanced system \textit{scalability and reliability}: By offloading processing tasks from central cloud servers, edge deployments may handle an increased volume of data, effectively scaling computational capabilities horizontally. Moreover, by performing local processing, edge systems can tolerate temporary network outages or disconnections to the cloud, maintaining operational continuity for critical tasks. This distributed resilience improves the overall reliability of the system, particularly in remote or intermittently connected environments.

\textbf{Devices at the Edge.}
Edge computing environments utilize a diverse array of devices, characterized by their heterogeneity in terms of computational capabilities. These devices range from highly resource-constrained entities, such as embedded microcontrollers and single-board computers (e.g., Raspberry Pi), to more powerful local servers or mini-data centers. An edge node typically comprises hardware components for computing (e.g., CPUs, GPUs), storage (e.g., solid-state drives, volatile memory), and network connectivity. Many modern edge devices also incorporate specialized accelerators, such as GPUs for parallel processing or Software-Defined Networking (SDN) components for optimized network routing and management, to enhance their performance for specific tasks.
Edge computing is frequently viewed as an extension of cloud computing, forming a layered architecture known as the edge-cloud continuum~\cite{AlDulaimy2024}. This continuum typically comprises distinct layers based on geographical proximity to data sources and computational capabilities:
The Edge Layer consists of devices directly at the periphery of the network, such as sensors, mobile devices, switches, and routers. These are the closest to the data generation. The Fog Computing Layer is positioned between the edge and the centralized cloud, involving intermediary nodes like servers or smaller data centers in closer geographical proximity to edge devices than the remote cloud. These nodes offer more substantial computing and storage capabilities than individual edge devices. The Cloud Layer is the uppermost layer. It is characterized by large, centralized data centers with extensive and highly scalable computational resources, with immense processing power. 
The edge-cloud continuum is not limited to hierarchical algorithms but also peer-to-peer models~\cite{Karagiannis2019}, facilitating direct inter-device communication

\begin{figure}[h]
\captionsetup{justification=centering}
\centering
    \begin{subfigure}{.3\textwidth}
        \centering
        \includegraphics[width=.9\textwidth]{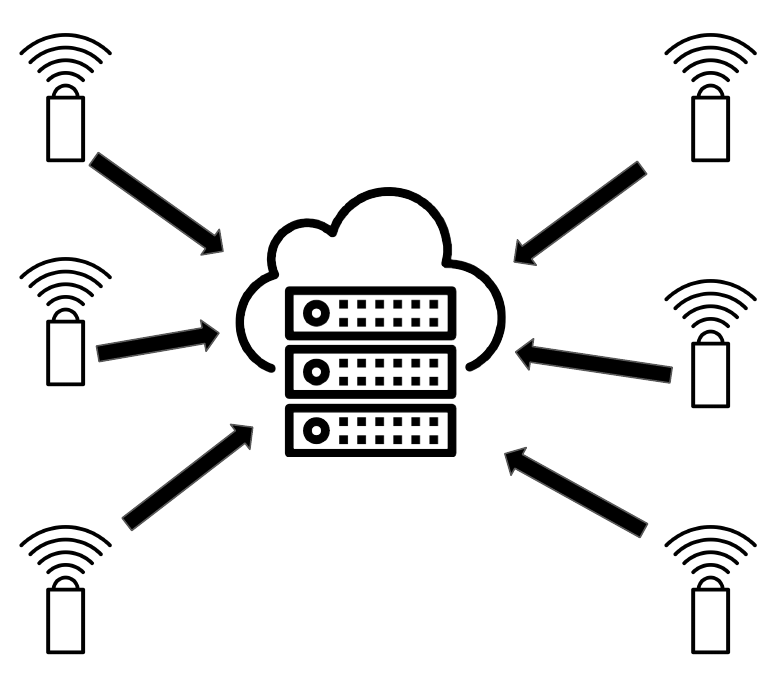}
        \caption{Mobile Cloud \\ Computing}
        \label{fig:mcc}
    \end{subfigure}
    \begin{subfigure}{0.3\textwidth}
        \centering
        \includegraphics[width=.9\textwidth]{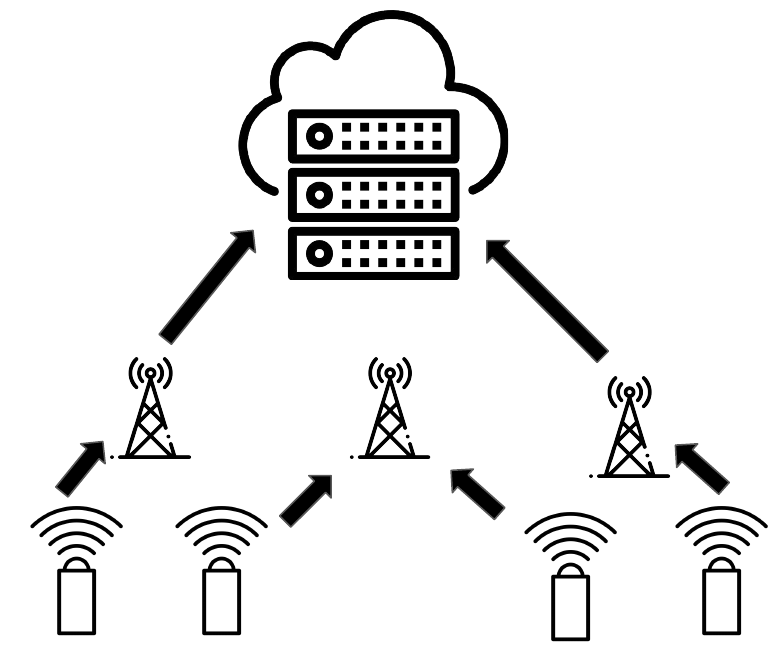}
        \caption{Multi-Access Edge \\ Computing}
        \label{fig:mec}
    \end{subfigure}
    \begin{subfigure}{.3\textwidth}
        \centering
        \includegraphics[width=.9\textwidth]{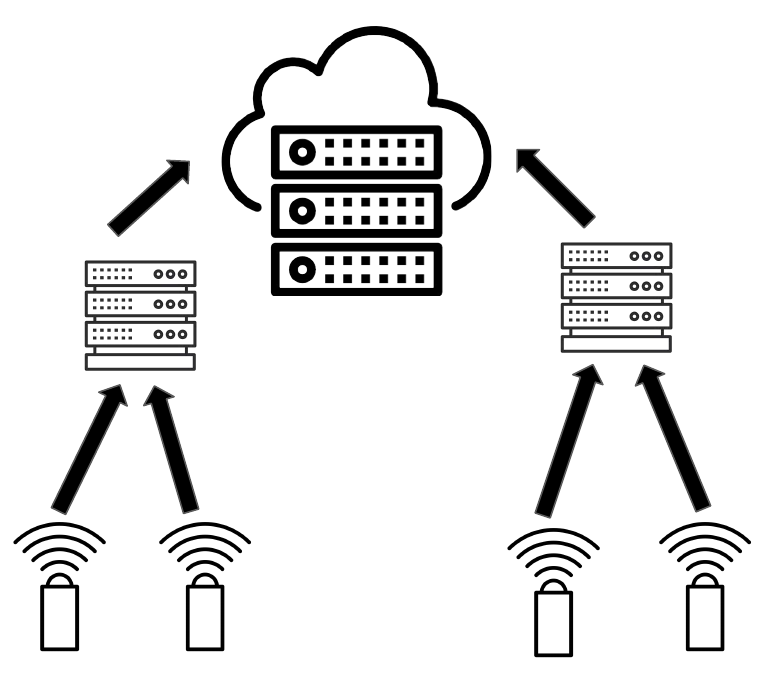}
        \caption{Cloudlet \\ Computing}
        \label{fig:cloudlet}
    \end{subfigure}
    \caption[short]{The three process mining topologies differ in computing and communication allocation, with Distributed PM leveraging data sources' inherent resources.}
    \label{fig:topologies}
\end{figure}

Within the edge-cloud continuum, various topological instantiations~\cite{Mahmud2017,AlDulaimy2024} exist and each topology comes with specific characteristics:
\textit{Mobile Cloud Computing (MCC)} as shown in Figure~\ref{fig:mcc} refers to mobile devices or sensors offloading computationally intensive operations to remote cloud servers by transmitting relevant data. The cloud acts as an extension for resource-constrained mobile devices.
\textit{Multi-Access Edge Computing (MEC, Figure~\ref{fig:mec})} primarily concerns mobile devices and sensors connecting to the network via access points (e.g., cellular base stations, Wi-Fi routers) that are equipped with computing, storage, and networking capabilities. MEC servers are deployed at the base station level or aggregation points to provide cloud-like services with low latency.
\textit{Cloudlets}(Figure~\ref{fig:cloudlet}) are small-scale, decentralized data centers that provide cloud-like computing and storage capabilities at the network edge, often in close physical proximity to mobile users or IoT devices. Cloudlets represent a form of fog computing.
This paper's evaluation employs all three topologies in the context of conformance checking. 

\textbf{Quality Attributes at the Edge.}
Understanding the elementary quality attributes is essential for designing and evaluating edge computing systems. According to a structured mapping study by Ashouri et al.~\cite{Ashouri2021}, the most prominent quality measures in edge computing are time behavior, functional suitability, resource utilization, and load capacity. While other attributes, such as reliability, security, and maintainability, also play a role, they appear with comparatively lower frequency in the surveyed literature.

\textit{Time Behavior} encompasses various metrics related to the timeliness of operations, including latency (the delay between an event and a system's response), throughput (the rate at which data is processed or transmitted), and response time (the total time taken for a system to respond to a request). Real-time requirements in many IIoT applications make time behavior a paramount concern.
\textit{Functional Suitability} refers to the degree to which a system provides functions that meet stated and implied needs when used under specified conditions. In edge computing, this often relates to the ability of edge devices to perform specific tasks accurately. In the realm of process mining, functional suitability could also refer to the integration of metrics such as fitness and precision~\cite{vanderAalst2022}.
\textit{Resource Utilization} quantifies how efficiently computing, storage, or network components are being used in proportion to their maximal possible load. Efficient resource utilization is crucial for optimizing the cost and performance of edge deployments, especially given the potentially constrained nature of edge devices.
\textit{Load Capacity} relates to the maximum workload or data volume a system can handle before its performance degrades unacceptably or fails. For edge systems, understanding load capacity is vital for ensuring that the distributed infrastructure can accommodate peak loads and maintain required service levels.

\section{Requirements for an Edge Process Mining Formalization}
\label{sec:epm}

The characteristics of edge computing environments necessitate a reevaluation of traditional process mining algorithm design principles. Classical process mining tools and methodologies are developed under the assumption of centralized data storage and computation. Deploying process mining algorithms at the edge introduces new design challenges. These implications primarily concern the distributed nature of data and computational resources, the need to incorporate edge computing quality measures, and the influence of varying hardware topologies. This section presents requirements to extend the current process mining formalism to edge computing.

\textbf{R1) Distributed Nature of Data.}
Traditionally, process mining algorithms operate on event data aggregated and stored in a central event log. In contrast, the distributed nature of data in edge computing environments implies that event data originates from multiple, geographically dispersed sources. This necessitates a shift towards a new notion of \textit{distributed event logs}, conceptually aligned with principles found in federated process mining~\cite{FederatedPM}. Furthermore, given the real-time requirements, this concept must be extended to \textit{distributed event streams}, where events are processed continuously as they are generated across the network.

\textbf{R2) Distributed Nature of Algorithms.}
Classical process mining algorithms typically assume complete and instantaneous access to all relevant event data. However, in an edge computing deployment, an algorithm initially has access only to the data originating from or localized within its own host edge device. To process data from other distributed devices or to gain a global perspective of a process, explicit inter-device communication is required. This means that algorithms must be designed to initiate or respond to network requests for data exchange. This distributed operational model introduces network communication between the edge nodes.

\textbf{R3) Modelling of a Heterogeneous Compute Topology.}
Unlike classical process mining, which typically presumes computation on a single machine with uniform access to data, the heterogeneous compute topology of edge environments becomes relevant. The performance of a process mining algorithm is influenced by where it is executed within the edge-cloud continuum, whether on a resource-constrained sensor, an intermediate fog node, or a centralized cloud server. Consequently, a formal notion for describing network requests and the underlying heterogeneous computing topology is essential. Such a formalization would enable systematic analysis and comparison of different deployment strategies and data distribution patterns across the edge-cloud continuum, allowing for optimized allocation of computational tasks.

\textbf{R4) Integration of Edge Computing Quality Measures.}
The deployment of process mining algorithms at the edge introduces new requirements concerning their performance and resource consumption. Algorithms must be designed to operate effectively under real-time constraints, ensuring timely insights and actions. Furthermore, they must adhere to the load capacity limitations of the underlying infrastructure. Therefore, the design of edge process mining algorithms must consider and optimize for edge computing quality attributes, such as low latency, efficient resource utilization, and high load capacities and scalability.
\section{PM-EdgeMap}
\label{sec:formal}

This section introduces a set of formal definitions intended to describe process mining algorithms within an edge computing context. The primary focus of this formalism is on modeling and analyzing real-time requirements. While other important concerns, such as data privacy, long-term persistence, or energy efficiency, are acknowledged, they are considered secondary within the scope of this specific formalization. The presented framework specifically models key quality measures, namely processing time, resource utilization, load capacity, and scalability. The overarching design goal is to provide a comprehensive and intuitive means to compare the performance of different process mining topologies and algorithms in distributed edge environments.
It is important to state the delimitations of this formalism. It does not aim to provide a fully detailed simulation of the intricate dynamics of an edge environment, which would necessitate the employment of complex modeling techniques such as queuing theory~\cite{adan2002queueing}, thereby significantly increasing model complexity. Furthermore, this formalism does not explicitly account for overutilization scenarios where existing queues might buffer tasks, but rather focuses on peak load capacities.

\textbf{Distributed Event Stream.}
An Event $e = (c, a, t, l)$ is a tuple of case $c \in \mathcal{C}$, activity $a \in \mathcal{A}$, timestamp $t \in \mathcal{T}$ and a location $l \in \mathcal{L}$. 
Thereby, the location indicates the data source (e.g., a sensor) from which the event originates. 
Let $\mathcal{E}$ denote the set of all possible events defined as $\mathcal{E} := \mathcal{C}\times \mathcal{A}\times \mathcal{T}\times \mathcal{L}.$ The function $\pi_l(e)$ returns the location of the event $e$. 
An Event Stream $S$ is a function from the natural numbers to an event $e \in \mathcal{E}$, i.e., $S:= \mathbb{N} \rightarrow \mathcal{E}$, with $S(i)$ being the i-th element of the event stream. The event stream $S_l$ located at $l \in \mathcal{L}$ is defined as a subset of the event stream $S$, where all events are located at $l$, i.e., $S_l := \{e |  i \in \mathbb{N}, e = S(i), \pi_l(e) = l\}.$ We refer to the union of all Located Event Streams as a Distributed Data Stream: $S_D := \bigcup_{l \in \mathcal{L}} S_l$. Further, we define the velocity of an event stream as the number of events emitted in a certain period of time $\Delta t$ as a function $v_{\Delta t}: S\rightarrow \mathbb{N}$.

\textbf{Computing Topology.}
We denote $N$ as the set of available computing nodes and $HC$ as the set of hardware components $hc = (node, t_{init}, r_{init}, tp, type)$ as a tuple of the $node \in N$ where the hardware component belongs to, the initial processing delay $t_{init}$, the initial resource usage $r_{init}$, and the throughput $tp$ and a type $type \in \{storage, compute, network\}$. To access a property from $hc$ a function with the property name is defined, e.g. to get the throughput the function $tp: HC \rightarrow \mathbb{R}$ exists.
An \textit{edge topology} $\mathcal{T} \subseteq \mathcal{P}(HC)$ is subsequently defined as the set of all interconnected edge nodes within the system.
A \textit{hardware instruction} $hi \in HI \subseteq HC\times\mathbb{R}$ denotes a tuple of a hardware component and a payload size value. These hardware instructions are used when translating the pseudocode of an algorithm to the set of invoked functions. Therefore, we define an \textit{algorithm execution} $\mathcal{A}: E\rightarrow \mathcal{P}(HI)$ for a single event $E$ as the ordered sequence of all hardware instructions that are made by the algorithm while processing that event. Further, we describe the function $payload: HI\rightarrow \mathbb{R}$ and the function $hw: HI\rightarrow HC$, which return the payload and the hardware component of a hardware instruction.

\textbf{Quality Measures.}
Based on the defined formalism, we model the edge quality attributes \textit{processing time}($t_{process}$), \textit{resource utilization}($r$), as well as the load capacity and scalability:
For a given event $e\in E$ and an algorithm $\mathcal{A}$, the \textit{processing time} $t_{process}(\mathcal{A},e)$ is defined as the sum of the initial delays and the processing times for all hardware calls within the execution path. Assuming sequential execution, the processing time is given by:

$$t_{process}(\mathcal{A}, e) = \sum_{step \in (\mathcal{A}(e))} \left( t_{init}(hw(step)) + \frac{payload(step)}{tp(hw(step))} \right)$$

The resource utilization $r$ is determined per hardware component $HC$ over a time unit $\Delta t$. For an event stream $S$ with velocity $v(S)$, and considering all hardware calls $hi$ for events $E$ within $\Delta t$ that utilize hardware components. Hence, the utilization of a hardware component is given by:

$$r(S,\mathcal{A}, hc)= \sum_{i \in \mathbb{N}}\,\sum_{\substack{step \in \mathcal{A}(S(i)) \\ hw(step) = hc}} \frac{v(S)\cdot(r_{init}(hw(step))+ payload(step))}{tp(hw(step))}$$


The \textbf{Load Capacity} of a system is defined via a service level objective (SLO). An exemplary SLO is that the resource utilization should not exceed 100\% for any component, or that the response time should stay below 100 milliseconds.
The $capacity$ is then defined as the maximum event stream velocity $v_S$ without violating that SLO. The \textit{scalability} in this context, describes how the load capacity changes when the resource throughput of individual hardware components $hc \in HC$ is increased~\cite{Henning2022}. It quantifies the system's ability to improve performance as resources are added or enhanced, indicating how effectively the system can grow to handle higher event stream velocities.
\section{Discussion of an Edge Conformance Checking Prototype}
\label{sec:algo}

To provide a concrete illustration of how process mining can be adapted for edge computing environments, this section outlines a specialized conformance checking algorithm designed to adhere to the requirements for edge process mining algorithms outlined in the preceding discussion. Conformance checking was selected as the focus for this implementation for several key reasons. Firstly, it directly reflects potential real-time requirements, particularly when applied in scenarios such as anomaly detection within a smart factory. In such contexts, deviations from an expected process model (i.e., a drop in conformance below a defined threshold) may necessitate immediate actions, such as halting production, thereby demanding stringent real-time analysis. Secondly, not all process mining tasks inherently require real-time processing; a process analyst might tolerate a certain degree of data outdatedness, for instance, up to one minute, for tasks like periodic performance monitoring. Conformance checking, however, can critically inform immediate operational decisions.

\begin{figure}[ht]
\captionsetup{justification=centering}
\centering
\includegraphics[width=\textwidth]{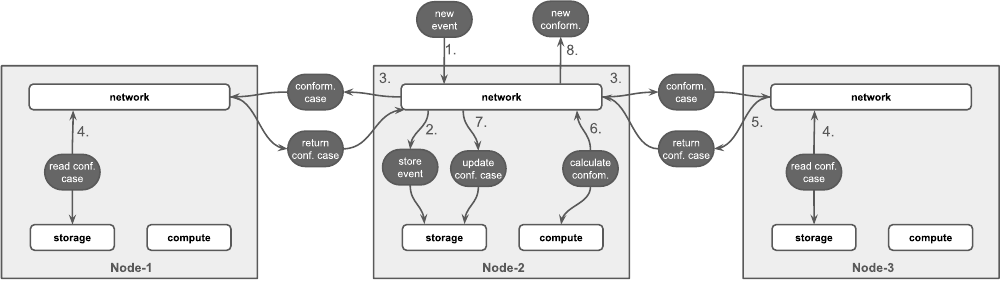}
\caption{Sketch of the algorithm execution for checking the conformance for an incoming event. The invoked node sends requests to other participants to check whether they observed events with the same case id and request their conformance.}
\label{fig:algo}
\end{figure}

The design of this edge conformance checking algorithm draws inspiration from existing work, specifically from the federated conformance checking algorithms proposed by~\cite{Rafiei2025} and the streaming conformance checking approaches developed by~\cite{Burattin2018}. The algorithm operates through a two-phase design: a training phase and a conformance phase. The training phase is dedicated to recording and analyzing normal process behavior within the smart factory use case, from which a DFG is constructed. Subsequently, the conformance phase continuously checks the conformance of currently running process instances against the learned model. Fig.~\ref{fig:algo}  visualizes the algorithm's conformance checking phase.

During the training phase, a significant departure from traditional approaches is observed. Instead of mining a single, global process model, each instance of the algorithm, typically residing on a specific edge node, learns and maintains its own localized DFG. When a new event is processed at a particular edge node, the algorithm must identify its predecessor within the associated case, which may have been observed on a different node. This necessitates a distributed lookup: a network request is initiated to all available nodes to ascertain whether they have observed an event belonging to the same case. The event with the timestamp closest to, but preceding the currently processed event, is then identified as the predecessor and subsequently integrated into the local DFG. This distributed learning mechanism enables the algorithm to adapt to localized process variations while contributing to a collective understanding of a process's flow.

The Conformance Phase is responsible for continuously tracking and evaluating the conformance for each active process instance. To achieve this, the current conformance state of every case is maintained in memory on the respective edge nodes. When an event represents an "inbound activity" (i.e., its predecessor was observed on a different edge node), a network request is dispatched to the predecessor node. This request queries the current conformance status of the running trace on that remote node, allowing the algorithm to aggregate conformance information across distributed segments of the process instance. This mechanism ensures that even for cases spanning multiple edge devices, a real-time assessment of conformance can be maintained.


The algorithm is implemented in Python together with a simulation of the PM-EdgeMap formalism. Both are publicly available on GitHub\footnote {https://github.com/cau-se/PM-EdgeMap/}. The evaluation has been run on the IoT data set presented in~\cite{Malburg22}. In this smart factory scenario, the machine type divides the events into a distributed event log, and the Distributed Event Factory disseminates the data stream~\cite{reiter2024}. We evaluated three different edge topologies: mobile edge computing, mobile cloud computing, and cloudlet. Table~\ref{tab:evaluation} holds the setup and the evaluation results.

\begingroup
\renewcommand{\arraystretch}{1.5} 
\begin{table}[]
    \centering  
    \caption{The setup of the edge computing topologies together with the corresponding quality metrics, processing time, and load capacity. The tuple for the hardware components describes the initial latency per request, the initially consumed resources, and the throughput of the component.}
    \begin{tabular}{|>{\centering\arraybackslash}m{2.3cm}|>{\centering\arraybackslash}m{3cm}|>{\centering\arraybackslash}m{3cm}|>{\centering\arraybackslash}m{3cm}|} 
    \hline 
    & \textbf{MCC} & \textbf{MEC} & \textbf{Cloudlet} \\
    \cline{1-4} 
    \noalign{\hrule height 1.2pt} 
    $network_{data}$    & $(0.025s, 2, 1000\frac{1}{s})$ & $(0.001s, 1, 250\frac{1}{s})$ & 
    $(0.025s, 1.5, 500\frac{1}{s})$ \\
    $network_{control}$ & $(0.005s, 1, 1000\frac{1}{s})$ & $(0.025s, 2, 250\frac{1}{s})$ & 
    $(0.025s, 1.5, 500\frac{1}{s})$ \\
    $cpu$               & $(10^{-5}s, 1, 10^4\frac{1}{s})$ & $(10^{-4}s, 1, 10^{4}\frac{1}{s})$ &
    $(2\cdot10^{-5} s, 1, 5\cdot 10^4\frac{1}{s})$ \\
    $storage$           & $(10^{-4}s, 1, 10^{4}\frac{1}{s})$ & $(10^{-3}s, 1, 10^{3}\frac{1}{s})$ &
    $(2\cdot10^{-4} s, 1, 5\cdot 10^3\frac{1}{s})$ \\
    \cline{1-4} 
    \noalign{\hrule height 1.2pt} 
    \
    $t_{process}$ & 0.036s & 0.043s & 0.027s \\
    $capacity$ & $83 \frac{events}{s}$ & $138 \frac{events}{s}$ & $174 \frac{events}{s}$ \\
    \hline 
    \end{tabular}
    \label{tab:evaluation}
\end{table}
\endgroup

In the MCC topology, all sensors are connected directly via a central data network. Conversely, in the MEC and Cloudlet topologies, data sources are distributed among four edge nodes, with each node connected to between two and four individual data sources. The specification of the nodes can be retrieved from Table \ref{tab:evaluation}.
The evaluation yielded distinct performance characteristics across the assessed topologies. Processing time was highest in MEC environments (0.043 seconds per event), while Cloudlet-based architectures demonstrated the lowest processing times (0.027 s/event). Cloudlet deployments exhibited the highest capacity (174 events/s), whereas MCC showed the lowest(83 events/s). The primary limiting factor for load capacity has been the data network in most scenarios. An exception is MEC, where the control network was the bottleneck. Another insight is that CPU resource consumption was predominantly consumed in the MCC scenario. We attribute this heightened consumption to the increased complexity of conformance checking alignments resulting from a larger data basis. Due to the performant CPUs in the modeled cloud, that fact did not influence the overall performance.

The PM-EdgeMap formalization and its prototype edge conformance checking algorithm are validated through fulfilling four key requirements. This includes the distributed processing of an IIoT datastream across computing locations (R1), the algorithm's efficient operation on data subsets (R2), a comparative analysis of edge computing topologies (R3), and the evaluation of processing time and load capacity as quality measures (R4). 

\section{Future Research Directions and Improvements}
\label{sec:dis}

Within this formalism's application, several threats to validity are identified, concurrently highlighting avenues for future work. For one, the performance of edge computing quality attributes is highly dependent on the assumed edge computing topology. Currently, processing times are predominantly limited by data network bandwidth, with compute and storage capabilities playing a minor role. This dynamic would fundamentally change with datasets comprising unprocessed, unstructured data, such as video or sensor streams, which are more resource-intensive for storage and demand greater compute for tasks like machine learning inference. Consequently, realistic evaluations necessitate execution on actual edge hardware, but given the limited access to edge testbeds, extending analysis to edge computing simulators~\cite{Jha2020} without increasing system complexity becomes crucial.

Furthermore, the presented edge conformance checking algorithm serves as a foundational example, illustrating core principles of distributed process mining. Its current simplification lacks advanced capabilities like loop or parallelism detection, vital for modeling complex real-world processes. Future designs must extend this framework to incorporate such functionalities. Additionally, the reliance on artificial data for evaluation limits external validity, emphasizing the need for comprehensive, representative distributed event logs from real IIoT scenarios or the development of an advanced event log generator. This would allow precise control over input parameters, including event velocity and data heterogeneity. Moreover, IIoT process mining often involves unstructured data, where preliminary pre-processing for event and case identification introduces additional computational overhead that must be accurately integrated into performance models. Lastly, rigorous validation ideally requires real hardware testing. Therefore, employing realistic simulation environments that emulate distributed, heterogeneous, and resource-constrained edge deployments is imperative. 
The integration of data streaming middleware would provide more realistic insights into performance, resource consumption, and scalability under operational IIoT conditions, thereby enhancing the external validity of research findings.

\section{Related Work}
\label{sec:rel}

This paper lays the groundwork for a new research field: process mining within the edge-cloud computing continuum. Because this area is so new, closely related works are rare. However, there is relevant research in distributed process mining, real-time process mining on data streams, and applications of IoT data processing on the edge-cloud-continuum.
Distributed process mining can be subdivided into two main types: distribution of data, like in federated process mining~\cite{FederatedPM,FederatedPM2}, and distribution algorithms. Federated process mining differs from our approach as it analyzes static event logs without considering real-time needs or the computing infrastructure. When algorithms are distributed, techniques such as MapReduce~\cite{MapReduce,FHM} are often used to make calculations faster and more efficient. Unlike our approach, these methods do not work on logically distributed data. Instead, they split data from a single event log for computations. While algorithms like EdgeMiner~\cite{Andersen} and CheckMyFlow~\cite{Andersen2025} do operate on distributed data and perform computations in a distributed manner, they focus on creating footprint matrices in a distributed way. Hence, they rather focus on preprocessing than on directly distributing the process mining algorithms.
In the area of real-time process mining~\cite{Burattin2022}, algorithms such as streaming process mining are used for process discovery and conformance checking. However, unlike our approach, these methods are not applied to distributed data.
Other data processing communities, such as machine learning, have already established paradigms for distributing their algorithms across the edge-cloud continuum for IoT data processing~\cite{Arzovs2024}. This is exemplified by techniques such as federated learning~\cite{Li2020} or TinyML~\cite{Zaidi2022}.
While Federated Learning enables distributed model training on decentralized datasets while preserving privacy, TinyML focuses on reducing the computational complexity to allow execution on resource-constrained devices. These fields of research may be used as inspiration for further development of process mining algorithms on the edge-cloud continuum.


\section{Conclusion}
\label{sec:con}

In this paper, we explored the synergistic integration of process mining and edge computing within the context of large-scale IIoT data. We highlighted how edge computing offers significant advantages by enabling low-latency and low-bandwidth data processing, while simultaneously fostering privacy-aware and scalable data handling. Our investigation systematically identified key implications for the design of process mining algorithms when deployed at the edge. These implications encompass the inherently distributed nature of event data and the algorithms themselves, along with the necessity to incorporate the underlying computing topology into the formal framework.

Building upon these insights, we derived a set of critical quality attributes specifically tailored for edge process mining, including processing time, resource utilization, load capacity, and scalability. To demonstrate the feasibility and practical applicability of our proposed formalism, we presented a case study centered on a real-time conformance checking algorithm. This demonstration effectively showcased how variations in the underlying edge-cloud topology, as well as adjustments to algorithmic parameters, directly influence performance within such distributed environments.

Looking ahead, our future work intends to advance the concept of edge process mining further. We plan to develop a more sophisticated edge conformance checking algorithm that explicitly incorporates advanced features such as parallelism and loop detection, crucial for handling the complexities of real-world industrial processes. Furthermore, we aim to enhance the validity and robustness of our design by conducting rigorous benchmarks on actual edge hardware, moving beyond simulated environments. Concurrently, our efforts focus on testing these algorithms with authentic distributed IIoT datasets to ensure their practical efficacy. Ultimately, our overarching goal is to enable the extraction of actionable, real-time process insights from the vast and intricate datasets characteristic of modern IIoT landscapes.

\begin{credits}
\subsubsection{\ackname} This work received funding from the Deutsche Forschungsgemeinschaft (DFG), grant 496119880
\end{credits}

%
%
%
\bibliographystyle{splncs04}
\bibliography{bibliography}
\end{document}